\documentclass{PoS}

\pdfoutput=1

\usepackage{upgreek}
\usepackage{wrapfig}
\usepackage{eurosym}

\title{The SST-1M camera for the Cherenkov Telescope Array}

\ShortTitle{CTA SST-1M camera}

\author{{\speaker{E.~J.~Schioppa}$^a$}
F.~Cadoux$^{a}$,
A.~Christov$^{a}$,
D.~della Volpe$^{a}$,
Y.~Favre$^{a}$,
M.~Heller$^{a}$,
T.~Montaruli$^{a}$,
A.~Porcelli$^{a}$,
M.~Rameez$^{a}$,
I.~Troyano Pujadas$^{a}$, 
W.~Bilnik$^{k}$,
J. B\l{}ocki$^{c}$,
L.~.Bogacz$^{m}$,
T~.Bulik$^{d}$,
M.~Cury{\l}o$^{c}$,
M.~Dyrda$^{c}$,
A.~Frankowski$^{g}$,
\L{}. Grudniki$^{c}$,
M.~Grudzi{\'n}ska$^{d}$,
B.~Id{\'z}kowski$^{e}$,
M.~Jamrozy$^{e}$,
M.~Janiak$^{g}$,
J.~Kasperek$^{k}$,
K.~Lalik$^{k}$,
E.~Lyard$^{b}$,
E.~Mach$^{c}$,
D.~Mandat$^{l}$,
A.~Marsza{\l}ek$^{c,e}$,
J.~Micha{\l}owski$^{c}$,
R.~Moderski$^{g}$,
A.~Neronov$^{b}$,
J.~Niemiec$^{c}$,
M.~Ostrowski$^{e}$,
P.~Pa{\'s}ko$^{f}$,
M.~Pech$^{l}$,
E.~Prandini$^{b}$,
P.~Rajda$^{k}$,
P.~Schovanek$^{l}$,
K.~Seweryn$^{f}$, 
K.~Skowron$^{c}$,
V.~Sliusar$^{j}$,
M.~Sowi{\'n}ski$^{c}$,
{\L}.~Stawarz$^{e}$,
M.~Stodulska$^{e}$,
M.~Stodulski$^{c}$,
S.~Toscano$^{b,n}$, 
R.~Walter$^{b}$,
M.~Wi{\c e}cek$^{k}$,
A.~Zagda\'{n}ski$^{e}$,
K.~Zi{\c e}tara$^{e}$,
P.~{\.Z}ychowski$^{c}$ (the SST-1M sub-consortium) for the CTA consortium\footnote{Full consortium author list at http://cta-observatory.org}.\\
\footnotesize{
a. D\'epartment de physique nucleaire et corpusculaire, Universit\'e de Gen\`eve, 
CH-1205 Switzerland.\\
b. ISDC, Observatoire de Gen\`eve, Universit\'e de Gen\`eve, 
1290 Versoix, Switzerland.\\
c. Instytut Fizyki J{\c a}drowej im. H. Niewodnicza{\'n}skiego Polskiej Akademii Nauk, 
31-342 Krak{\'o}w, Poland.\\
d. Astronomical Observatory, University of Warsaw, Al. Ujazdowskie 4, 00-478 Warsaw, Poland\\
e. Astronomical Observatory, Jagiellonian University, ul. Orla 171, 30-244, Krakow, Poland.\\
f. Centrum Bada{\'n} Kosmicznych Polskiej Akademii Nauk,  18a Bartycka str., 00-716 Warsaw, Poland.\\
g. Nicolaus Copernicus Astronomical Center, Polish Academy of Sciences,  Warsaw, Poland.\\
j. Astronomical Observatory, Taras Shevchenko Nat. University of Kyiv, Observatorna str., 3, Kyiv, Ukraine.\\
k. AGH University of Science and Technology, al.Mickiewicza 30, Krakow, Poland,\\
l. Institute of Physics of the Czech Academy of Sciences, 
Prague, Czech Republic.\\
m. Department of Information Technologies, Jagiellonian University, 
30-348  Krakow, Poland.\\
n. Vrije Universiteit Brussels, Pleinlaan 2 1050 Brussels, Belgium.\\}
}

\abstract{The prototype camera of the single-mirror Small Size Telescopes (SST-1M) proposed for the Cherenkov Telescope Array (CTA) project has been designed to be very compact and to deliver high performance over thirty years of operation. 
\newline
The camera is composed of an hexagonal photo-detection plane made of custom designed large area hexagonal silicon photomultipliers and a high throughput, highly configurable, fully digital readout and trigger system (DigiCam). The camera will be installed on the telescope structure at the H. Niewodnicza{\'n}ski institute of Nuclear Physics in Krakow in fall 2015.
\newline
In this contribution, we review the steps that led to the development of the innovative photo-detection plane and readout electronics, and we describe the test and calibration strategy adopted.
}

\FullConference{The 34th International Cosmic Ray Conference,\\
		30 July- 6 August, 2015\\
		The Hague, The Netherlands}

\begin{document}

\section{Introduction}
The Cherenkov Telescope Array (CTA) is an array of ground-based telescopes proposed for studying gamma rays by detecting the UV-blue Cherenkov light generated in atmospheric showers. A sub-array of CTA will be composed of Small Size Telescopes (SST) and will be dedicated to the high energy region of the gamma spectrum, from 5 to 300~TeV. In order to acquire enough statistics at such high energies, the SST sub-array will need to cover a large (several km$^2$) surface on the ground. Key features for the prototyping of the SST telescopes are thus the low cost and the large scale producibility. 
\newline
The innovative prototype camera proposed for the single mirror Small Size Telescopes (SST-1M) has been conceived with these goals in mind. Among the novelties of the design, the sensor units composing the photo-detection plane are custom produced Geiger-Avalanche PhotoDiodes (G-APD, or Silicon Photomultipliers, SiPM). As demonstrated by the experience with the FACT telescope, a SiPM-based camera can be operated in half moon conditions, thus achieving a 30\% longer exposure time if compared to cameras based on Photomultipliers Tubes (PMT)~\cite{FACT}. Moreover, SiPMs do not undergo ageing due to light exposure, are more robust to high intensity light and the working point is more stable in the long term.

\section{Camera structure and mechanics}
\begin{wrapfigure}{r}{0.5\textwidth}
	\centering
	\includegraphics[width=0.5\textwidth]{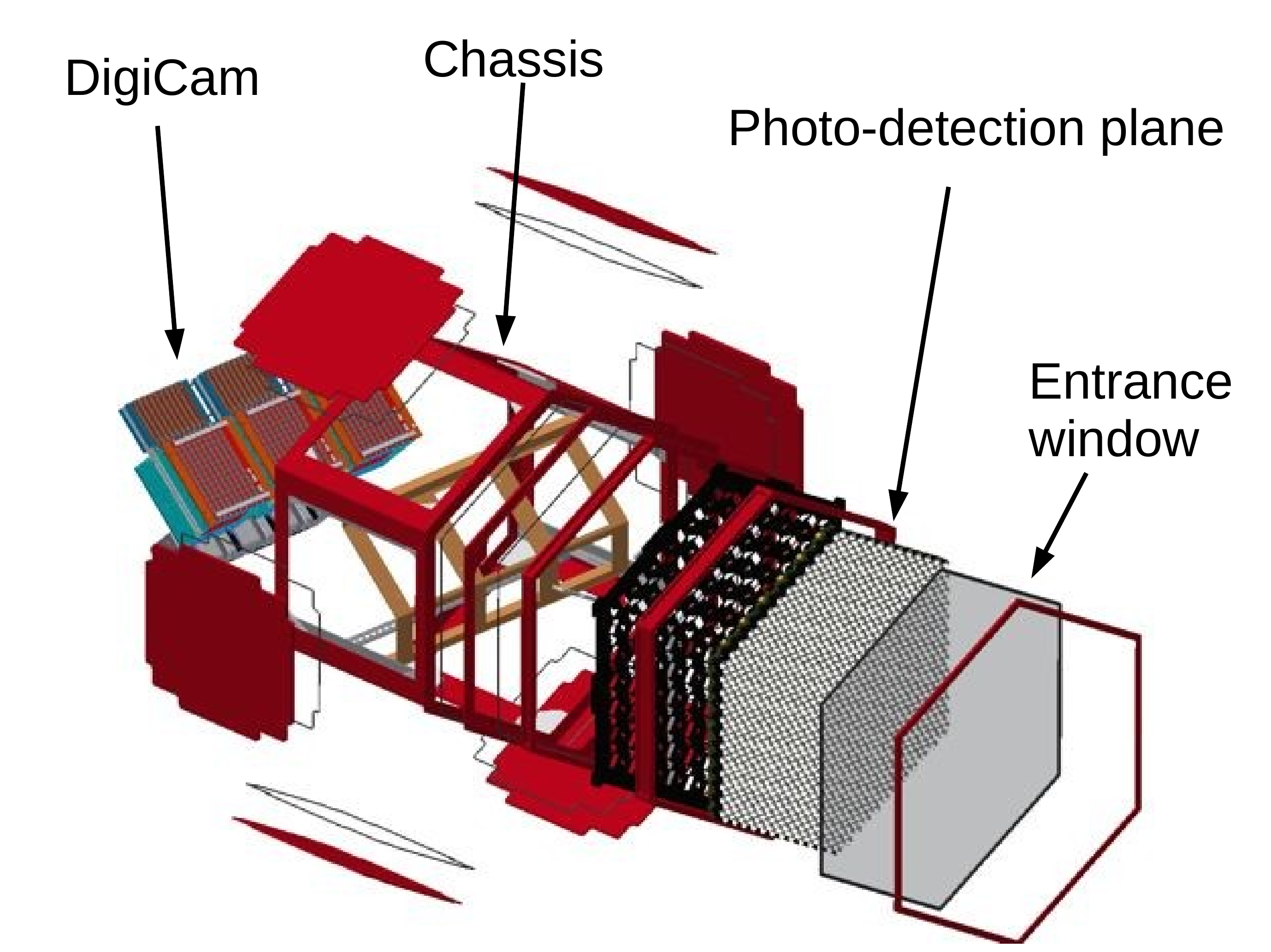}
	\caption{Exploded view CAD drawing of the SST-1M camera.}
	\label{camera CAD}
\end{wrapfigure}
A CAD drawing of the SST-1M camera is shown in figure~\ref{camera CAD}. An innovative feature of the camera is the physical separation between the Photo-Detection Plane (PDP) and the readout and trigger electronics (DigiCam). Both components are enclosed in an aluminium box, wit a 3.3~mm thick Borofloat window at the PDP side. The box is shaped into an hexagonal cylinder 90~cm flat-to-flat broad and 60~cm deep. A six petals - six motors shutter installed at the PDP side will ensure the light tightness of the camera at parking position and during dedicated calibration runs, while providing protection from the weather conditions. The camera protection from water end dust is achieved with and IP65 compliant design. The overall structure is very compact and lightweight (less than 200~kg).

\section{Photo-detection plane}
\begin{figure}
	\centering
	\includegraphics[width=0.4\textwidth]{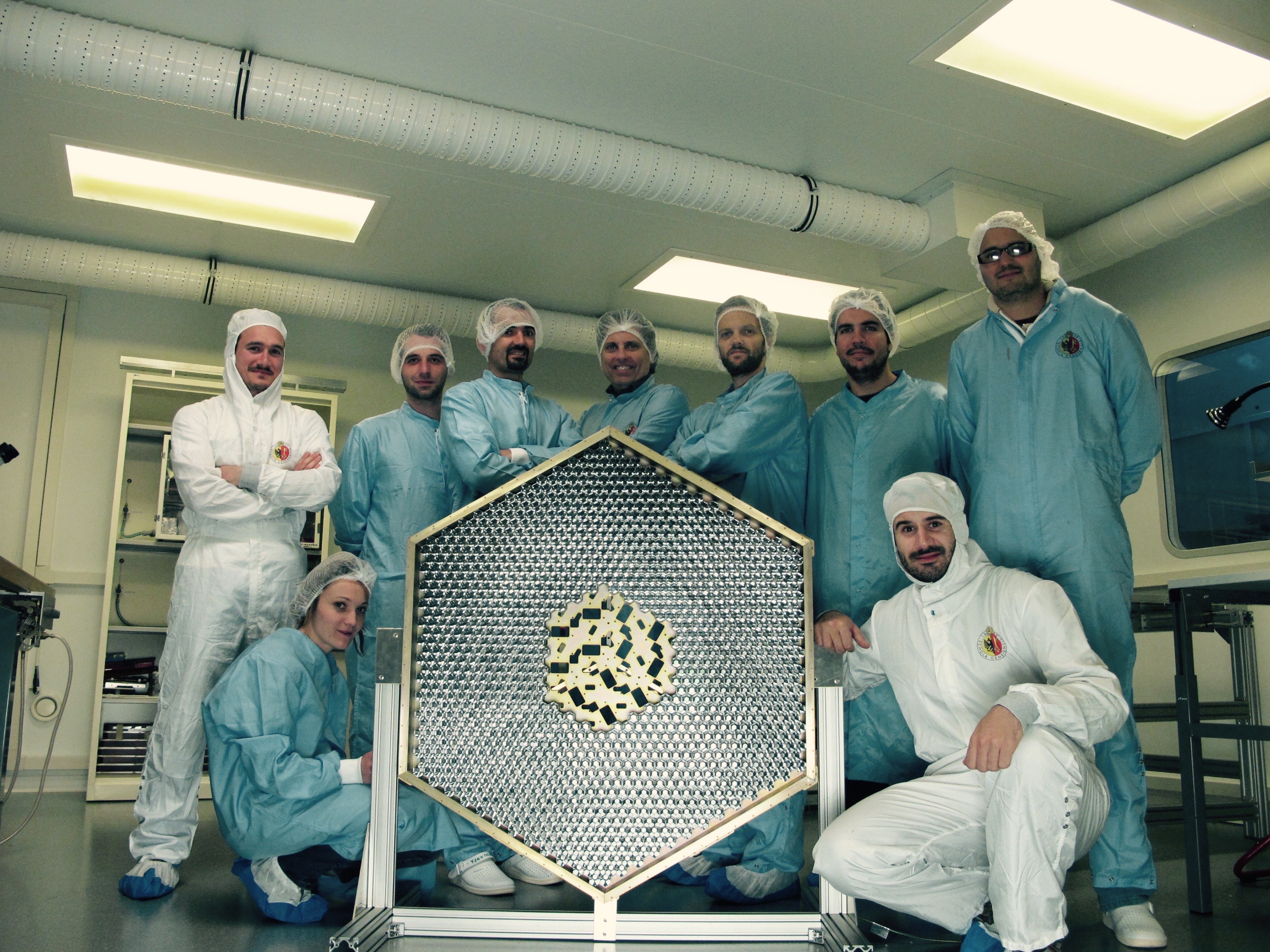}
	\qquad
	\includegraphics[width=0.26\textwidth]{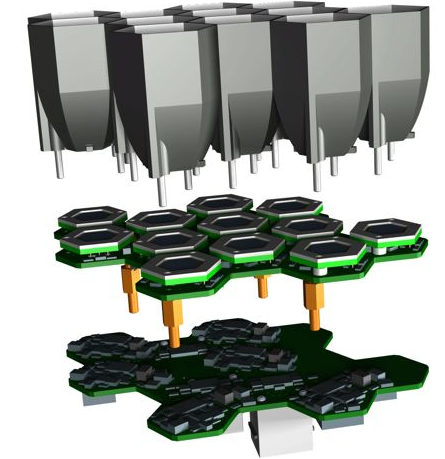}
	\caption{A picture of the PDP (left) assembled in the laboratories of the University of Geneva, except for its central part that was being used for the cooling validation tests the time of the picture, and a drawing of a single 12 pixels module (right) highlighting its main components: the Winston cones, the sensors attached to the PreAmp board and the Slow Control Board.}
	\label{module}
\end{figure}

The Photo-Detection Plane (PDP) is a two dimensional array of hexagonal sensor units (the pixels), that is being developed at the University of Geneva. The hexagonal pixel shape is crucial to achieve an optimal trigger performance. Such geometry guarantees that each pixel center is at the same distance from all its
neighbors, so no preferred direction exists and trigger algorithms can benefit from such a symmetry.
\newline
The 1296 pixels of the PDP are arranged into 108 individual modules (see figure~\ref{module}) of 12 pixels each, directly connected to the front-end electronics boards.
The modules are mounted on an aluminium backplate, that guarantees the mechanical stability of the PDP but also serves as an active part of the cooling system (see later on). The PDP is 88~cm broad flat-to-flat and weights 35~kg.

\subsection{Light concentrators}
\begin{figure}
	\centering
	\includegraphics[width=0.2\textwidth]{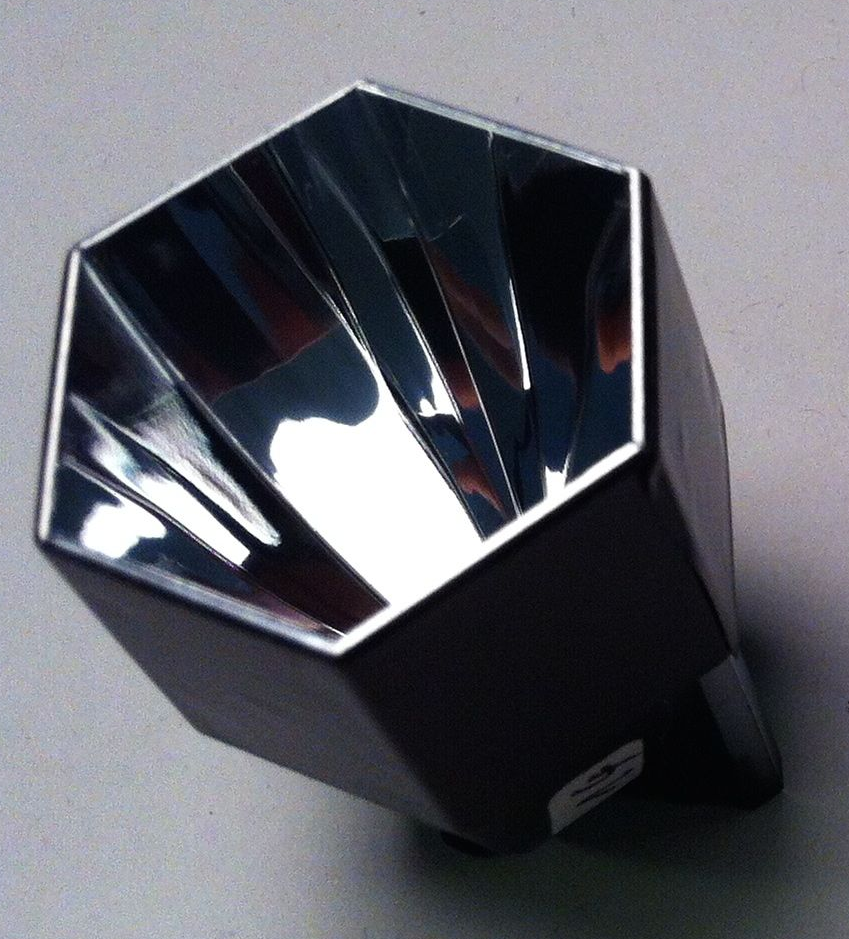}
	\includegraphics[width=0.3\textwidth]{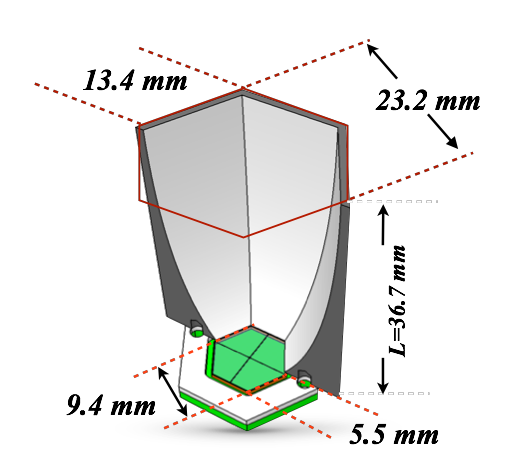}
	\includegraphics[width=0.45\textwidth]{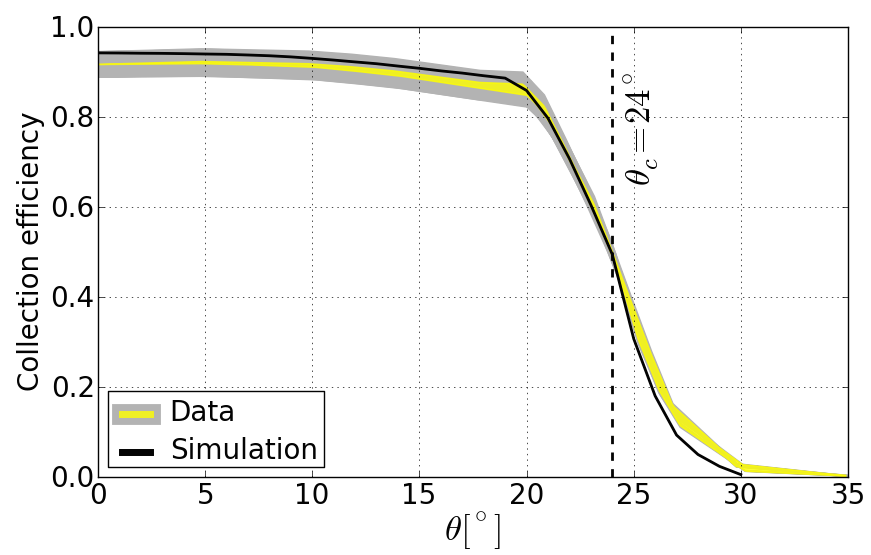}
	\caption{A picture of a single cone (left), a drawing highlighting its geometrical properties with respect to the underlying sensor (center) and the measured light transmission efficiency as a function of the angle of incidence $\theta$, showing the cut-off at 24$^\circ$ (right).}
	\label{cones}
\end{figure}
The required 9$^\circ$ field of view and few arcmin angular resolution of the SST camera imply, for single-mirror optics, a possible hexagonal pixel size of 2.32~cm flat-to-flat. Since no SiPM of such a linear dimension exists, the solution adopted for the SST-1M camera is to use concentrators to focus the light onto a sensor with smaller area. To match the hexagonal pixel surface to a standard SiPM with square surface, the only possibility is to use a plastic light guide. However, compressing the hexagonal pixel entrance into a standard 3~$\times$~3~mm$^2$ square sensor would require a quite long light guide (around 4-5~cm), which would absorb most of the UV light. To avoid this, the use of hollow Winston cones has been proposed~\cite{conesPaper}. The geometry of the cone is fixed by the pixel size and by the requirement of having a cut-off for light coming at angles above 24$^\circ$ (to reduce the stray light coming from directions outside the field of view and to maximize the collection efficiency of the light focused by the mirror dish on the photosensor), and has been optimized for the manufacturing process using Bezier curves to shape the six faces of the hexagonal cone (see figure~\ref{cones}). The cones are made of polycarbonate and are produced using the cheap injection molding technique. An aluminium and dichroic layers coating covers the inner surface to enhance the UV reflectivity at large angles of incidence.

\subsection{SiPM sensors}
\begin{wrapfigure}{r}{0.5\textwidth}
	\centering
	\includegraphics[width=0.15\textwidth]{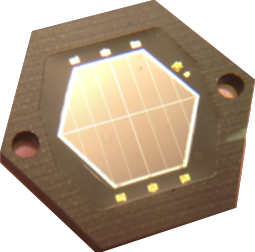}
	\qquad
	\includegraphics[width=0.22\textwidth]{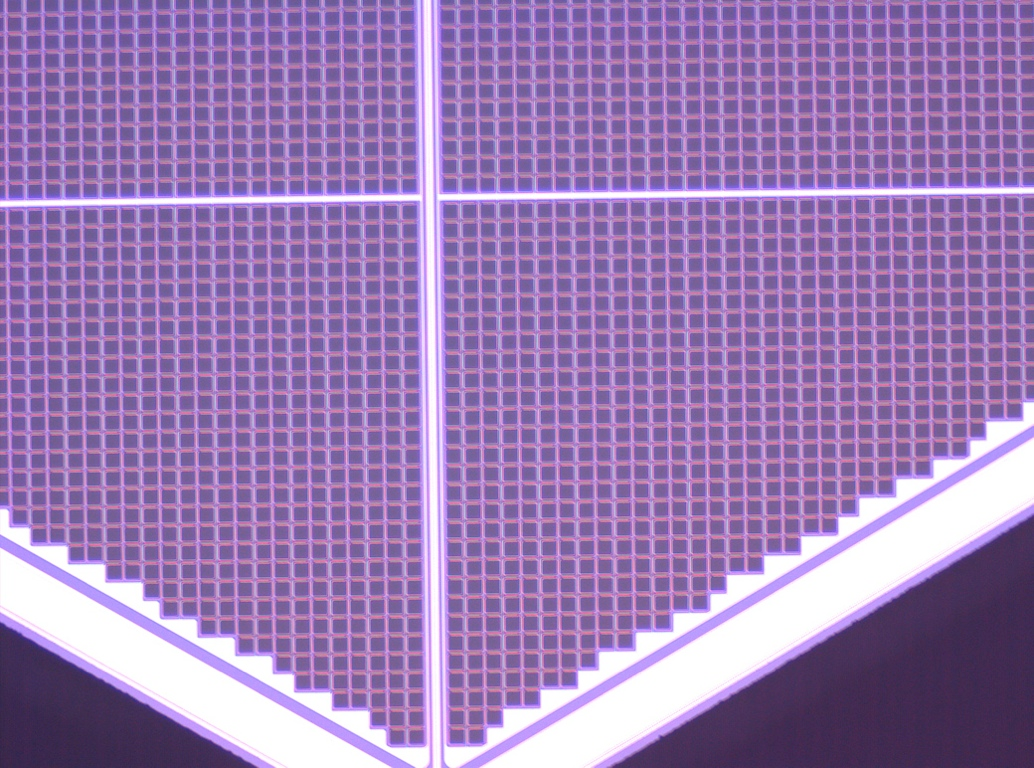}
	\caption{A picture of the large area hexagonal SiPM (left) and a microscope image showing individual microcells (right).}
	\label{sensor}
\end{wrapfigure}
The geometry of the light concentrator determines the size and shape of the underlying sensor as 9.4~mm side-to-side on an hexagonal surface. Such large area SiPMs, with such a peculiar shape, are not commercially available, and have thus been designed for the purpose in a collaboration with Hamamatsu exploiting standard manufacturing techniques for low cross-talk SiPMs~\cite{MPPC}. A picture of the sensor, together with a zoom on its 50~$\upmu$m size square cells, is shown in figure~\ref{sensor}.
\newline
The high number of microcells (36840 square cells in total, each of 50~$\upmu$m size) that can be packed on the large sensor area (93.56~mm$^2$) ensures the operation of the device far from saturation. However, the large area produces a high capacitance (3.4~nF), that translates into a long duration of the signals (in the order to 100~ns), thus limiting the bandwidth capability of the device to below the required 10~MHz rates. To overcome this limitation, the 36840 cells have been split in four channels (9210 cells each, 850~pF capacitance) in common cathode configuration.

\subsection{Front-end electronics}
\label{front-end section}
Due to space constraints, the front-end electronics of each module is implemented in two separate PCBs, the preamplifier board (PreAmp) and the Slow Control Board (SCB). The design of the two boards has been driven by the need of having a low noise, high-bandwidth, low power and limited cost front-end electronics.

\subsubsection{PreAmp board}
\begin{wrapfigure}{r}{0.5\textwidth}
	\centering
	\includegraphics[width=0.5\textwidth]{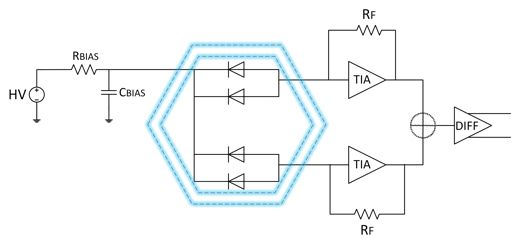}
	\caption{Schematics of the preamplifier.}
	\label{front-end}
\end{wrapfigure}

The PreAmp board amplifies and shapes the signals coming from the sensors. In order to provide a single readout channel per pixel, the signals coming from the four channels of a single sensor are recombined via the summation circuitry shown in figure~\ref{front-end}. Here transimpedence stages are used to amplify the charge, so that the resulting signals bear information about the amount of detected light.
\newline
An innovative feature is the DC coupling of the sensors to the front-end, which will allow to monitor the baseline position as a function of the background light level, giving access to direct information about the night sky background (NSB). The NSB is foreseen to be order of 30~MHz per pixel in dark nights, considerably higher than the dark noise rate of about 5-10~MHz.

\subsubsection{Slow Control Board}
The Slow Control Board (SCB) is a key component of the PDP. Its main functions are to provide the bias voltage to each sensor of the module individually and to route the analog signals from the PDP to DigiCam via standard RJ45 connectors. Furthermore, the SCB implements a complex logic that ensures the stability of the working point of the sensors. In fact, the gain of a SiPM is determined by the overvoltage, i.e. the difference between the bias voltage and the breakdown voltage, which is strongly dependent on temperature. In fact, average gradients in the order of 5~$^\circ$C are expected over the PDP surface. The design of an active cooling system to stabilize the temperature would be challenging. The adopted solution is to realize a logic that, at a frequency of 2~Hz, stabilizes the working point of each sensor individually as a function of temperature. The temperature is measured from a negative temperature coefficient (NTC) probe that has been implemented in the sensor on purpose, with a precision of 0.17~$^\circ$C. The compensation loop is implemented in a microcontroller integrated in the SCB, and it also features a real-time control logic that, using a periodic readout of the bias voltage, checks and compensates for rounding errors. The bias voltage is adjusted with a precision of 6.69~mV.
\newline
The SCB is accessible via CANbus, which also allows to directly read the temperature and read/write the bias voltage.

\section{DigiCam}
\begin{figure}
	\centering
	\includegraphics[width=1\textwidth]{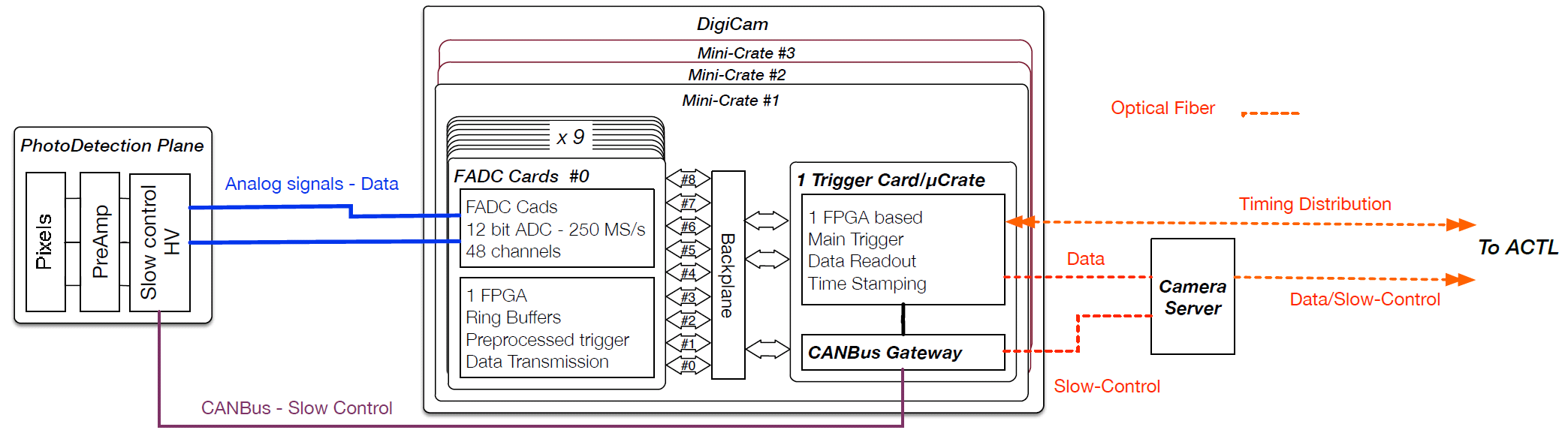}
	\caption{A schematics of the readout architecture.}
	\label{DigiCam}
\end{figure}
The approach of having a fully digital, dead time free, readout and trigger system, DigiCam, is another innovative feature of the camera, allowing for high flexibility in the implementation of the readout and trigger logic. The DigiCam hardware has been developed by the Jagiellonian University in Kr\`akow, and consists of 3 minicrates located at the back of the camera, each containing 9 digitizer boards and one trigger board interconnected via a custom backplane. For the readout and trigger purposes, the PDP is, in fact, logically divided into three sectors, each handled by one of the minicrates. All the data processing taking place in DigiCam is done by high throughput, latest generation, FPGAs (Xilinx XC7VX415T) present on each board. The digitizer board hosts 12 fast ADCs (4-channel AD9239 converters from Analog Devices, Inc.) for the digitization of the signals coming from 48 pixels, at the sampling rate of 250~MHz. The digitized signals are stored in ring buffers, and are processed to generate a L0 trigger decision before being sent to the trigger board located in the same minicrate. Of the three trigger boards, one is configured as master and the other two as slaves. The trigger boards are in charge of collecting the data on DDR3 memories, of calculating the L1 trigger decision and of sending the data to the central acquisition system via the master trigger card on a 10Gb fiber ethernet link. Within each crate, a highly parallelized trigger algorithm is applied to the data of the corresponding sector plus the neighboring channels of the adjacent sectors. The memories integrated on the trigger boards can be expanded in order to cope with the required rates. More information about DigiCam can be found in~\cite{DigiCam}. A schematics of the full readout architecture is shown in figure~\ref{DigiCam}.

\section{Cooling}
\begin{wrapfigure}{r}{0.4\textwidth}
	\centering
	\includegraphics[width=0.4\textwidth]{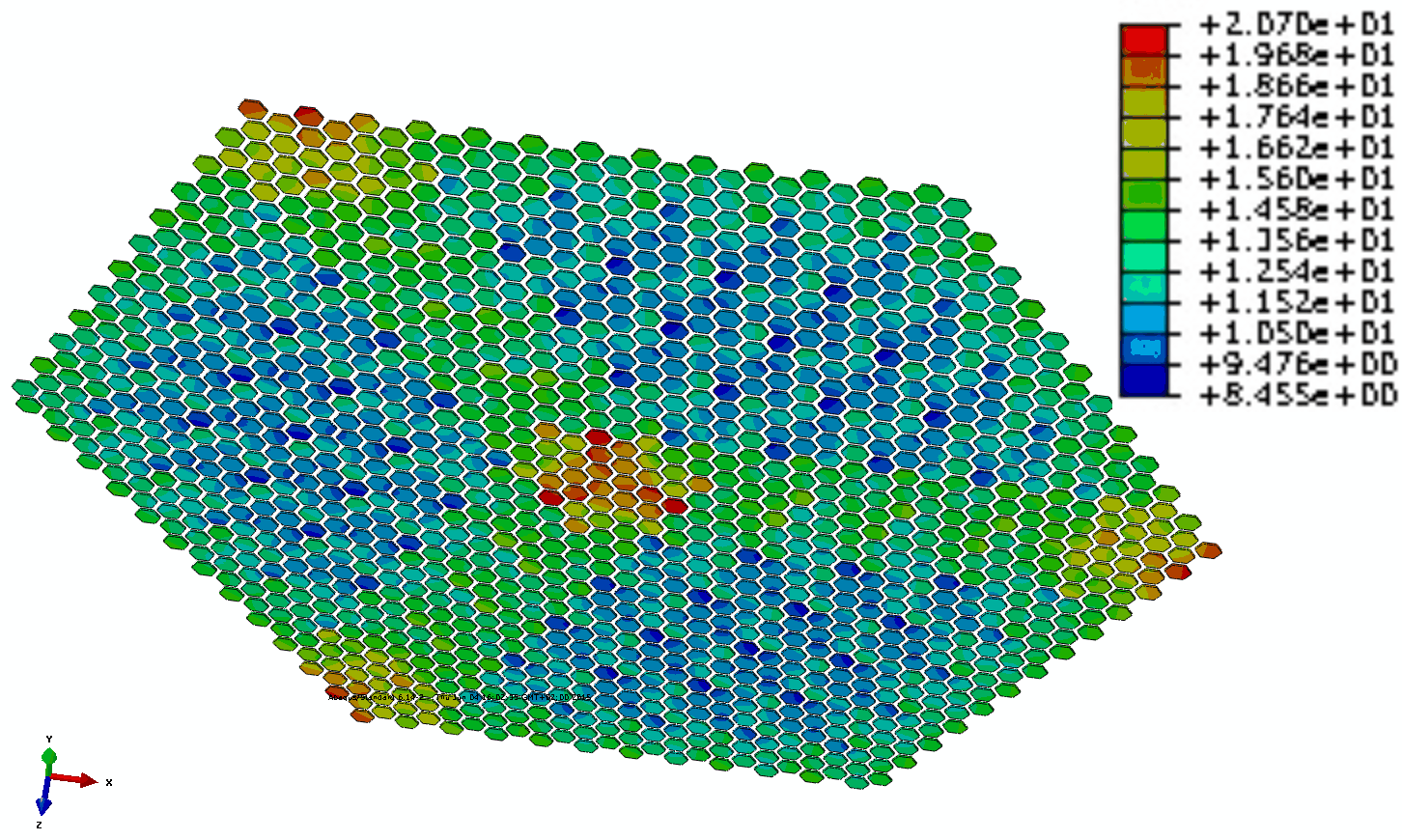}
	\caption{FEM calculation of the PDP temperature when the cooling system uses water at 7~$^\circ$C.}
	\label{cooling test}
\end{wrapfigure}
The total power consumption of the camera is $\sim$2~kW, where the contributions of the PDP and of DigiCam are 500~W and 1.2~kW, respectively. The cooling system was designed to comply with the IP65 insulation requirement on the camera structure and to ensure an efficient heat extraction from the very compact electronics.
\newline
The cooling of the PDP is done via a liquid coolant system. Cold water (at 10~$^\circ$C or less, chilled by external cooling units) flows through aluminium pipes fixed directly onto the backplate of the PDP, which thus serves as a heat sink, via aluminium blocks. The four screws that secure each module to the backplate serve as cold fingers to extract the heat from the front-end electronics. In order to ease the heat extraction from both the PreAmp and the SCB and to minimize temperature gradients, the two boards have been produced with a thicker copper layer (72~$\upmu$m instead of the conventional 18~$\upmu$m). Moreover, a thermally conductive filler material is inserted between the boards and between the fully assembled module and the backplate. The PDP cooling concept has been validated on a 1:10 mock-up of the PDP and on FEM calculations (see figure~\ref{cooling test}).
\newline
The heat is extracted from the DigiCam boards using heat pipes coupled with the water cooling pipes via a metal block as exchanger. To guarantee the proper functioning of the heat pipes at any inclination of the telescope, the DigiCam minicrates are mounted at a 45$^\circ$ angle, as shown back in figure~\ref{camera CAD}.

\begin{figure}[ht]%{r}{0.45\textwidth}
\centering
	\begin{minipage}[t]{0.38\textwidth}
%		\captionsetup{width=.99\textwidth} 
		\centering
		\includegraphics[width=1\textwidth]{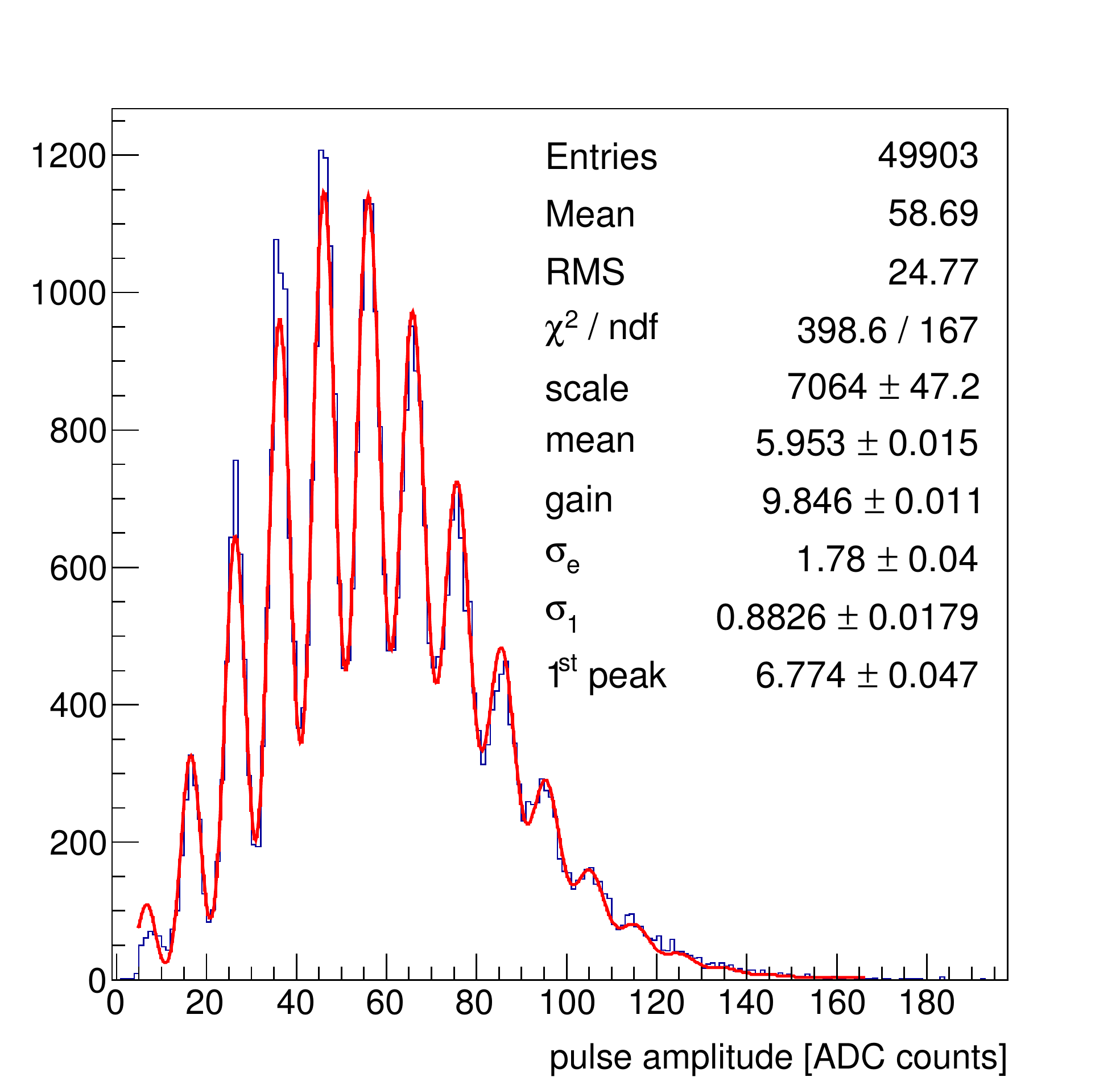}
		\caption{Single photon spectrum of a SiPM for an average of 10 photons illuminating the sensor.}
		\label{finger plot}
	\end{minipage}
	\begin{minipage}[t]{0.59\textwidth}
%		\captionsetup{width=.85\textwidth} 
		\centering
		\includegraphics[trim = 0mm 6mm 0mm 0mm, clip, width=1\textwidth]{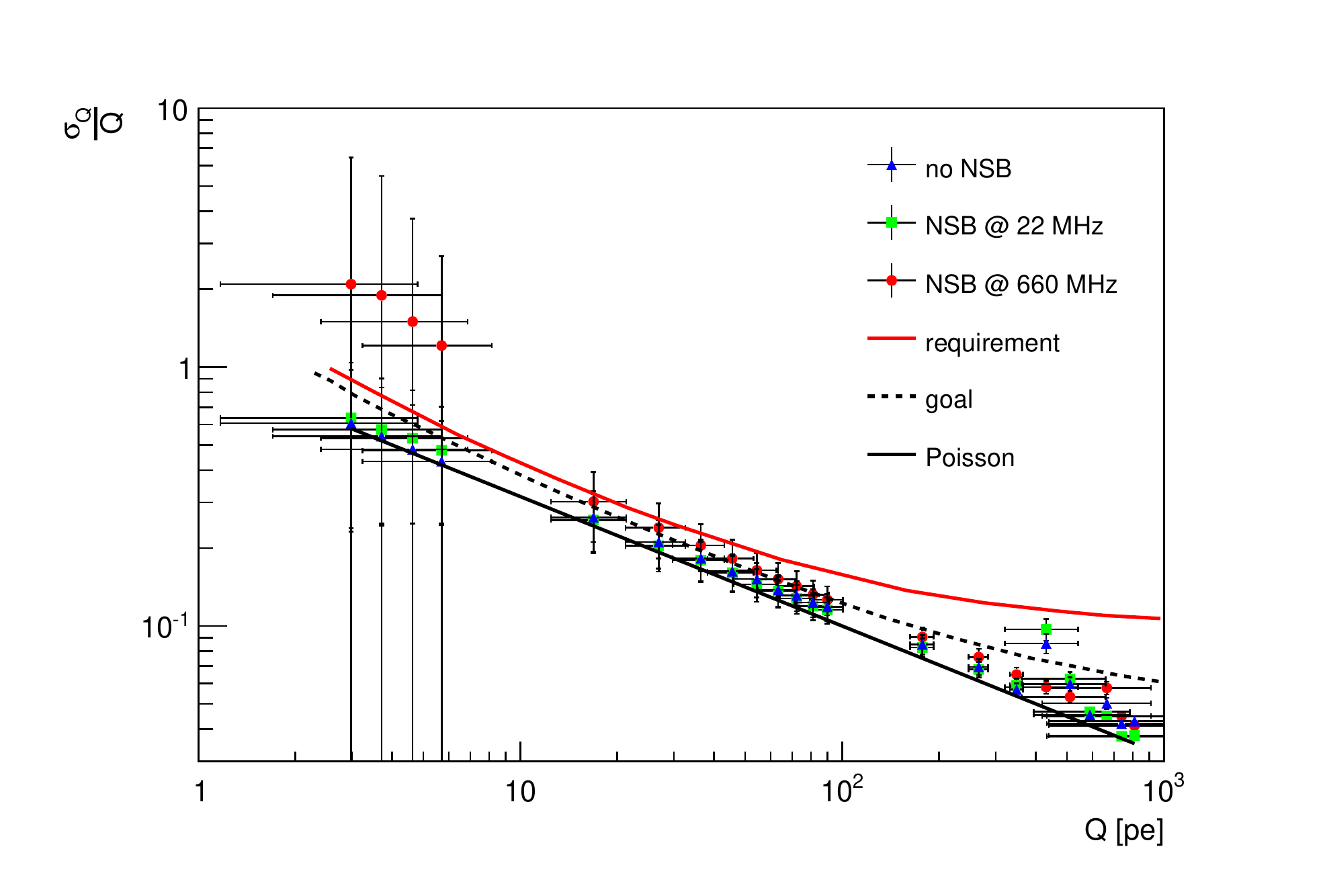}
		\caption{Charge resolution for a single sensor at different Night Sky Background (NSB) levels.}
		\label{charge resolution}
	\end{minipage}
\end{figure}
\section{Production, quality check and assembly}
In view of a possible large scale production following the prototyping phase, quality check procedures have been established for the various components prior to the assembly of the camera. The quality of the cones has been assessed on a subset of the full batch. Due to the high uniformity measured, there is in fact no need for a one-by-one characterization. The SiPM sensors have been characterized and tested by Hamamatsu. Dedicated electronics boards and control software have been realized for testing the PreAmp and the SCB at the production factory, which, for the case of the SCB, perform a preliminary calibration. Each setup has been designed to be highly portable and easy to operate. The results of the PreAmp and SCB tests have revealed a very good uniformity of the production.
\newline
The assembly of the single 12 pixels modules has been performed at the University of Geneva. Before being fixed on its final location on the PDP, each module undergoes an optical test using a dedicated setup. In this system, four modules can be illuminated simultaneously by the light emitted from a 470~nm LED source and transmitted up to each pixel through optical fibers. Using both pulsed light and continuous light, the setup allows to characterize the behavior of the modules in terms of the main signal parameters (e.g. baseline position, peak amplitude, rise time and fall time) for each sensor, as well as ADC gain uniformity, performance of the compensation loop, and so on. These data are used to calculate the flat fielding coefficients of the module, and can be also used in the future in the analysis of the actual data from the detected gamma-ray events (see also~\cite{calibration strategy}). The level of electronic cross talk has also been measured by illuminating one pixel in a module and by looking at the signals seen by all the other pixels. The result of the test shows that induced signals are only visible at high light levels (orders of 3000~photons), and even in this case the effect is in the order of few per-mill.

\section{Performance}
Preliminary studies on the performance of the camera detector units have been carried out. Single sensors or full modules have been tested in terms of optical properties using LED sources. Figure~\ref{finger plot} shows as an example the single photon spectrum obtained from one pixel of one of the assembled modules. The separation between the single photo-peaks is good despite the four channels of the SiPM are biased with an average operational voltage, as a consequence of the common cathode configuration. Figure~\ref{charge resolution} shows the charge resolution measured from one sensor with a relatively large gain spread between channels. Despite this, the charge resolution for dark night conditions (22~MHz NSB) falls below the goal line for almost all the charge range, and even the data points for half moon conditions (660~MHz NSB) are below requirements for light levels above 10 photo-electrons, and below the goal above 60 photo-electrons.

\section{Conclusion}
With its innovative features, the camera prototype for the SST-1M telescope proposed for the Cherenkov Telescope Array is foreseen to provide a high performance in the detection of air showers induced by high energy gamma rays in the atmosphere. While the camera is being assembled and commissioned at the University of Geneva, each of its components (light concentrators, sensors, electronics, and so on) is being tested and checked, and all the related data is being stored. This will provide a database of all the relevant operational parameters on a single-pixel basis, that will be helpful in the analysis of the actual data to ensure that the full potential of the camera can be exploited.

\section{Acknowledgements}
We gratefully acknowledge support from the agencies and organizations listed under Funding Agencies at this website: http://www.cta-observatory.org/. In particular we are grateful for support from the NCN grant DEC-2011/01/M/ST9/01891 and the MNiSW grant 498/1/FNiTP/FNiTP/2010 in Poland and the University of Geneva and the Swiss National Foundation.

\end{document}